\documentstyle[12pt,amsfonts]{article}
\newlength{\abstractwidth}
\renewcommand{\thefootnote}{\fnsymbol{footnote}}
\renewcommand{\thanks}[1]{\footnote{#1}} 
\newcommand{\starttext}{ \setcounter{footnote}{0}
\renewcommand{\thefootnote}{\arabic{footnote}}}

\newcommand{\be}{\begin{equation}}
\newcommand{\bea}{\begin{eqnarray}}
\newcommand{\eea}{\end{eqnarray}} \newcommand{\ee}{\end{equation}}
 
 \def\ba{\begin{eqnarray}}
\def\ea{\end{eqnarray}}

\def\cL{{\cal L}}

\def\cS{{\cal S}}

\def\log{\,{\rm log}\,}

\def\ge{\geq}
\def\le{\leq}

\def\R{{\bf R}}

\def\p{\partial}

\def\cD{{\cal D}}
\def\cF{{\cal F}}

\def\[{{\bf [}}
\def\]{{\bf ]}}

\def\ric{\textrm{Ric}}
\def\vol{dvol}
\def\Cl{\textrm{Cl}}

\def\cP{\mathcal{P}}
\def\cS{\mathcal{S}}

\begin{document}
\starttext \baselineskip=18pt \setcounter{footnote}{0}
\newtheorem{theorem}{Theorem}
\newtheorem{lemma}{Lemma}
\newtheorem{corollary}{Corollary}
\newtheorem{definition}{Definition}
\newtheorem{conjecture}{Conjecture}
\newtheorem{proposition}{Proposition}

\begin{center}
{\Large \bf A GEOMETRIC CONSTRUCTION OF SOLUTIONS
TO 11D SUPERGRAVITY\footnote{Work supported in part by National Science Foundation grants DMS-12-66033 and DMS-17-10500.}}
\nonumber \\
\medskip
\centerline{
Teng Fei, Bin Guo, and Duong H. Phong}

\bigskip

\begin{abstract}

\medskip

{\small Necessary and sufficient conditions are provided for a class of warped product manifolds with non-vanishing flux to be supersymmetric solutions of $11D$ supergravity. Many non-compact, but complete solutions can be obtained in this manner, including the multi-membrane solution initially found by Duff and Stelle. In a different direction, an explicit $5$-parameter moduli space of solutions to $11D$ supergravity is also constructed which can be viewed as non-supersymmetric deformations of the Duff-Stelle solution.}

\end{abstract}

\end{center}

\baselineskip=15pt

\section{Introduction}
\setcounter{equation}{0}

The $11D$ supergravity theory was first constructed by Cremmer, Julia, and Scherk \cite{cremmer1978}. The bosonic part of its action is given by
\be
\cL(g,{\mathcal A})=\int_M \frac{1}{2}R\, \vol- \frac{1}{4} F \wedge * F + \frac{1}{2} {\mathcal A} \wedge F \wedge F.
\ee
Here $g$ is a Lorentzian metric  on an oriented 11-dimensional manifold $M$ with one time-like direction, $R$ is the scalar curvature of $g$, $*$ is the Hodge star operator, ${\mathcal A}$ is a 3-form on $M$ and $F=d{\mathcal A}$ is the 4-form field strength (flux). The $11D$ supergravity theory occupies a privileged position in unification efforts including gravity, as the highest dimensional supergravity theory with no particle of spin greater than $2$, and as a low-energy limit of $M$ theory
(see e.g. \cite{T,W,HW, duff} and references therein). Its profoundly geometric nature makes its solutions not just interesting from the theoretical physics viewpoint, but also from the mathematics viewpoint, where they may ultimately serve as models of canonical metrics in new settings.

\medskip

It is well-known that the equations of motion of the theory, i.e., the critical point equation of the action $\cL$, are given by
\bea
\label{eqn:form} &&d*F=\frac{1}{2}F\wedge F,\\
\label{eqn:curv} &&\ric_{ij} = \frac 1 2 (F^2)_{ij}- \frac 1 6 |F|^2 g_{ij}.
\eea
In (\ref{eqn:curv}), $\ric_{ij}$ is the Ricci curvature tensor and $(F^2)_{ij}$ is the symmetric tensor given by
$$(F^2)_{ij}=\frac{1}{3!}F_{iklm}F_j^{~klm}=(\iota_{\p_i}F,\iota_{\p_j}F).$$
Here we follow the convention that for a $p$-form $F$ one writes
\bea
&&F=\frac{1}{p!}F_{i_1\dots i_p}d x^{i_1}\wedge\dots\wedge d x^{i_p},\nonumber\\
&&|F|^2=\frac{1}{p!}F_{i_1\dots i_p}F^{i_1\dots i_p}=\frac{1}{p!}F_{i_1\dots i_p}F_{j_1\dots j_p}g^{i_1j_1}\dots g^{i_pj_p}.\nonumber
\eea
The simplest solutions to $11D$ supergravity equation are those with trivial flux, i.e., $F=0$, in which case the equations reduce to the vacuum Einstein equation
$$\ric_{ij}=0.$$
Therefore we may think of the $11D$ supergravity equation as a generalization of the Einstein equation by turning on the 4-form flux $F$.

\medskip

Particularly interesting solutions of the $11D$ supergravity equation are the supersymmetric ones, i.e. solutions $(M^{11},g,F)$ to the equations of motion (\ref{eqn:form}) and (\ref{eqn:curv}) which admit a nonzero spinor $\xi$ satisfying
\be\label{eqn:susy}
\cD_m\xi:=\nabla_m\xi-\frac{1}{288}F_{abcd}(\Gamma^{abcd}_{~~~~m}+8\Gamma^{abc}\delta^d_{~m})\xi=0,
\ee
where $\nabla$ is the Levi-Civita connection induced on the spinor bundle and $\Gamma$ are Gamma-matrices acting as endomorphism of spinors. In other words, the Levi-Civita connection is twisted by the field strength $F$ to produce a connection $\cD$ on the spinor bundle, and supersymmetry requires the existence of a parallel spinor under the twisted connection $\cD$.

\medskip

Some well-known solutions of $11D$ supergravity are the following. Suppose the 11-dimensional Lorentzian manifold $(M^{11},g_{11})$ is a metric product of a Lorentzian 4-manifold $(M^4,g_4)$ and a Riemannian 7-manifold $(M^7,g_7)$ and $F=c\,\vol_4$ is a nonzero constant multiple of the volume form associated to $(M^4,g_4)$. Under such an assumption we see that (\ref{eqn:form}) is automatically satisfied. Moreover (\ref{eqn:curv}) reduces to two equations on $M^4$ and $M^7$ respectively:
\bea
(\ric_4)_{ij}=-\frac{c^2}{3}(g_4)_{ij},\qquad
(\ric_7)_{ij}=\frac{c^2}{6}(g_7)_{ij},\nonumber
\eea
i.e., $(M^4,g_4)$ and $(M^7,g_7)$ are Einstein manifolds with negative and positive scalar curvature respectively. This is the famous Freund-Rubin solution \cite{freund1980}, which includes compactifications of the form $AdS^4\times S^7$.

Shortly after Freund-Rubin's discovery, Englert \cite{englert1982} found that one can deform the Freund-Rubin solution on $AdS^4\times S^7$ by turning on flux on the $S^7$. Mathematically we can extend Englert's construction as follows.

Let us assume that $M^7$ has a $G_2$ structure with fundamental 3-form $\varphi$ and 4-form $\psi=*_7\varphi$. We will use the convention that $|\varphi|_7^2=|\psi|_7^2=7$. Assume that $F$ is of the form
$$F=c_4\vol_4+c_7\psi,$$
where $c_4$ and $c_7$ are constants. (\ref{eqn:form}) implies that
$$d\varphi=c_4\psi,$$
hence $(M^7,g_7)$ is a nearly $G_2$ manifold (see for example \cite{friedrich1997}), which implies that
$$(\ric_7)_{ij}=\frac 3 8 c_4^2(g_7)_{ij}.$$
It follows that (\ref{eqn:curv}) reduces to
\bea
(\ric_4)_{ij}=\frac{2c_4^2-7c_7^2}{6}g_{ij},
\qquad
\frac 3 8 c_4^2=\frac 1 6 (2c_7^2-c_4^2).\nonumber
\eea
Again the solutions are just products of Einstein manifolds with opposite signs of scalar curvature. Similarly the Pope-Warner solution \cite{PW, dW1} arises by making another choice of $F$ exploiting the structure of $M^7$. Many more methods have since been developed to find solutions, including other ansatz for the flux $F$, classifications by the number of supersymmetries preserved, by holonomy, and construction of Lax pairs. The literature on the subject is immense, and we can only refer here to a few representative papers \cite{H2, GP, DH, PvN, Fig}, in which more references can be found.

\smallskip

The main focus of the present paper will be rather on solutions of $11D$ supergravity which are warped products. Warped products are well-known mathematical constructions, but they appear to have been considered first in compactifications in string theory by de Wit et al \cite{dW2},
Hull \cite{H1}, and Strominger \cite{S}. An early application to solutions of $11D$ dimensional supergravity was by Duff and Stelle \cite{duff1991}, which will be of particular interest to us and will be discussed in greater detail in section \S 4.
More precisely we consider general warped products $M^{11}=M^3\times M^8$ as in (\ref{g}) below, with the ansatz (\ref{F}) for the flux $F$. We give necessary and sufficient conditions for such configurations to be a supersymmetric solution or just a solution of $11D$ supergravity (Theorem 2 and Theorem 3). The implementation of these conditions turns out to be surprisingly simple: in effect, it suffices to have a Ricci-flat manifold $\bar M_8$, equipped with a strictly positive harmonic function. While these two requirements combined exclude the possibility of a compact manifold $\bar M_8$ and a smooth harmonic function, they allow for a wealth of examples constructed from either a compact Ricci-flat manifold, or complete Ricci-flat manifolds with faster than quadratic volume growth (Theorem 5). In both cases, by results of Cheng-Li \cite{CL} and Li-Yau \cite{LY}, the Green's function is positive and can be used as the harmonic function. Remarkably, the construction of complete Ricci-flat manifolds with maximum volume growth is a topic of great current interest in mathematics, and the results obtained recently there, for example by Conlon and Rochon \cite{CR}, Li \cite{Li}, and Sz\'ekelyhidi \cite{Sz}
can be put to good use through Theorem 5 to produce new supersymmetric solutions of $11D$ supergravity. Finally we return with this new understanding to the Duff-Stelle solution. With the ansatz of Duff-Stelle, namely $\bar M_8$ is conformally flat and radially symmetric, it is easy to see that the explicit expressions obtained in \cite{duff1991} follow at once from Theorem 3. On the other hand, if we give up on the requirement of supersymmetry and try only to solve the field equations, we find not just the Duff-Stelle solution, but in fact a whole $5$-parameter family of solutions. It is an interesting mathematical problem to determine whether some analogues of Theorems 3 and 5 can hold in the absence of supersymmetry.

\section{Supersymmetry and field equations}
\setcounter{equation}{0}

The goal of this section is to classify all supersymmetric solutions to the $11D$ supergravity equation on $M^{11}=M^3\times M^8$ of the form
\bea
\label{g}
g_{11}&=&e^{2A}g_3+g_8, \\
\label{F}
F&=&d vol_3\wedge df,
\eea
where $g_3$ is a Lorentzian metric on $M^3$, $g_8$ a Riemannian metric on $M^8$, $d vol_3$ the volume form associated to $g_3$, $A$ and $f$ are smooth functions on $M^8$. It is convenient for us to refer to this geometric set-up just as $(g_3,g_8,A,f)$.

\medskip

Throughout this paper, we say that $(g_3,g_8,A,f)$ is a solution to $11D$ supergravity if $f$ is not a constant and the pair $(g_{11},F)$
solves the equations of motion (\ref{eqn:form}) and (\ref{eqn:curv}). If $f$ is a constant, then $F=0$ and the equations of motion reduce to the vacuum Einstein equation. Therefore we only consider the case where $f$ is not a constant.

\smallskip

We say that $(g_3,g_8,A,f)$ is supersymmetric if the pair $(g_{11},F)$
admits a nontrivial spinor, i.e. a section of the spin bundle ${\cS}_{11}$, called $\xi$, such that
\be\label{eqn:kspinor}
\cD_P\xi:= (\nabla_{11})_P\xi-\frac{1}{288}F_{QRST}\left(\Gamma^{QRST}_{~~~~~~P}+8\Gamma^{QRS}\delta^T_{~P}\right)\xi=0.
\ee
We say that $(g_3,g_8,A,f)$ is a supersymmetric solution if $(g_3,g_8,A,f)$ is a solution to $11D$ supergravity with supersymmetry.

\medskip

Throughout this section, as in (\ref{eqn:kspinor}),
we will use capital Latin letters $P,Q,R,S,T$ as indices for the 11-manifold $M^{11}=M^3\times M^8$, Greek letters $\alpha,\beta,\gamma$ as indices for $M^3$, and lowercase Latin letters $a,b,c,d$ as indices for $M^8$. The symbol $\nabla$ always denotes the Levi-Civita connection, whose subscript indicates the reference metric. For instance, $\nabla_3$ is the Levi-Civita connection with respect to the metric $g_3$.

\medskip

For any real vector space $V$ equipped with a quadratic form $q$, we define the associated Clifford algebra $\Cl(V,q)$ as
$$\Cl(V,q)=T(V)/I_q(V),$$
where $T(V)$ is the tensor algebra of $V$ and $I_q(V)$ is the ideal generated by $v\otimes v-q(v)$. If $q$ is a non-degenerate pairing of signature $(r,s)$, we will use the short hand notation $\Cl(r,s)$ for $\Cl(\R^{r+s},q)$.

As associative algebras, it is well-known that
$$\Cl(2,1)\cong\R(2)\oplus\R(2),\quad \Cl(8,0)\cong\R(16),$$
hence
$$\Cl(10,1)\cong\Cl(2,1)\otimes\Cl(8,0)\cong\R(32)\oplus\R(32).$$
Here we have denoted by $\R(m)$ the algebra of $m\times m$ real matrices.
Let $\{\gamma_\alpha\}_{\alpha=1,2,3}$ be the standard generator of $\Cl(2,1)$ and $\{\Sigma_a\}_{a=1,\dots,8}$ the standard generator of $\Cl(8,0)$. Write $\Sigma_9=\Sigma_1\Sigma_2\dots\Sigma_8$ which satisfies
$$\Sigma_9^2=1,\quad \Sigma_9\Sigma_j+\Sigma_j\Sigma_9=0,\quad\forall j=1,\dots,8.$$
Let $\{\Gamma_P\}_{P=1,\dots,11}$ be the standard generators of $\Cl(10,1)$. It is straightforward to check that an explicit isomorphism $\Cl(10,1)\cong \Cl(2,1)\otimes \Cl(8,0)$ is given by
$$\{\Gamma_P\}_{P=1,\dots,11}=\{\gamma_\alpha\otimes\Sigma_9,1\otimes\Sigma_a\}_{\alpha=1,2,3; ~a=1,\dots,8}.$$

In addition, we use $\cP$ to denote pinor representations, i.e., irreducible representations of Clifford algebras, and $\cS$ to denote spinor representations, i.e., irreducible representations of the even part of Clifford algebras. Same letters are used for pinor and spinor bundles over manifolds. From the structure results stated above, we know that $\Cl(2,1)$ has exactly two inequivalent pinor representations ${\cP}_3^\pm$ and both of them are 2-dimensional. When restricted to the even part $\Cl(2,1)^0$, both pinor representations are isomorphic to the spinor representation ${\cS}_3$. As for the Clifford algebra $\Cl(8,0)$, there is a unique pinor representation ${\cP}_8$ of dimension $16$, which decomposes as the direct sum of two inequivalent spinor representations:
$${\cP}_8={\cS}_8^+\oplus{\cS}_8^-,$$
where ${\cS}_8^\pm$ are the eigenspaces of $\Sigma_9$ with eigenvalue $\pm1$. Moreover, we have the following isomorphisms
\bea
{\cP}_{11}^\pm&\cong&{\cP}_3^\pm\otimes {\cP}_8,\nonumber\\
{\cS}_{11}&\cong& {\cS}_3\otimes {\cP}_8.\nonumber
\eea

 In order to classify all supersymmetric solutions, the first step is to pin down $g_3$.

\begin{lemma}
If $(g_3,g_8,A,f)$ is a solution to 11D supergravity, then $g_3$ has to be an Einstein metric.
\end{lemma}
\noindent{\it Proof.} Plugging in the ansatz into (\ref{eqn:form}), (\ref{eqn:curv}), the equations of motion reduce to
\bea
&&d(e^{-3A}*_8 df)=0,\label{eqn:harmonic}\\
&&(\ric_8)_{ab}-3(\nabla_8^2A)_{ab}- 3A_aA_b+\frac{e^{-6A}}{2}f_af_b-\frac{e^{-6A}}{6}|\nabla_8 f|^2(g_8)_{ab}=0,\label{eqn:8curve}\\
&&(\ric_3)_{\alpha\beta}=\left(e^{2A}(\Delta_8A+3|\nabla_8A|^2)-\frac{e^{-4A}}{3}|\nabla_8 f|^2\right)(g_3)_{\alpha\beta}.\label{eqn:3curv}
\eea
As $A$ and $f$ are independent of $M^3$, (\ref{eqn:3curv}) implies that there exists a constant $\lambda$ such that
$$\ric_3=\lambda g_3$$
and
\be
e^{2A}(\Delta_8A+3|\nabla_8A|^2)-\frac{e^{-4A}}{3}|\nabla_8 f|^2=\lambda\label{eqn:3curv2}.
\ee
Hence $g_3$ must be Einstein.

From now on, we will always assume that $g_3$ is Einstein with Einstein constant $\lambda$. The next step is to understand supersymmetry.

\begin{theorem}
Suppose $g_3$ is Einstein in the sense that $\ric_3=\lambda g_3$. Then the tuple $(g_3,g_8,A,f)$ is supersymmetric if and only if

{\rm (a)} $\lambda=0$,

{\rm (b)} $df=\pm d(e^{3A})$,

{\rm (c)} and the conformally changed metric $\bar{g}_8=e^Ag_8$ admits a covariantly constant spinor with respect to its Levi-Civita connection $\bar{\nabla}_8$.

\end{theorem}

\noindent{\it Proof.} To help analyze supersymmetry, we first consider the auxiliary product metric $g'_{11}=g_3+g_8$. Since ${\cS}_{11}\cong {\cS}_3\otimes {\cP}_8$ holds pointwise, we may identify the spinor bundle ${\cS}'_{11}$ associated to $g'_{11}$ as the tensor product of the spinor bundle ${\cS}_3$ of $g_3$ with the pinor bundle ${\cP}_8$ of $g_8$. In addition, there is an isometry of vector bundles $(TM^{11},g_{11})\cong(TM^{11},g'_{11})$ given by
$$(X_3,Y_8)\mapsto (e^AX_3,Y_8),$$
therefore we may further identify the spinor bundle ${\cS}_{11}$ associated to $g_{11}$ with ${\cS}_3\otimes {\cP}_8$ as well. Let $\{e_\alpha\}_{\alpha=1}^3$ be a local orthonormal frame of $g_3$ and $\{e_{a}\}_{a=1}^8$ a local orthonormal frame of $g_8$, then
$$\{e_P\}_{P=1}^{11}=\{e^{-A}e_\alpha,e_a\}_{\alpha=1,2,3,~a=1,\dots,8}$$
is a local orthonormal frame for $g_{11}$. Write $e_{\tilde{\alpha}}=e^{-A}e_\alpha$ and let $\epsilon$ be a local section of ${\cS}_3$ and $\eta$ a local section of ${\cP}_8$, then the Clifford multiplication associated to $g_{11}$ under above identification is given by
$$\Gamma_{\tilde{\alpha}}(\epsilon\otimes\eta)=(\gamma_\alpha\epsilon)\otimes(\Sigma_9\eta)$$
and
$$\Gamma_a(\epsilon\otimes\eta)=\epsilon\otimes(\Sigma_a\eta).$$
In addition, the Levi-Civita connection $\nabla=\nabla_{11}$ can also be identified as
\bea
&&\nabla_\alpha(\epsilon\otimes\eta)=((\nabla_3)_\alpha\epsilon)\otimes\eta +\frac{1}{2}(\gamma_\alpha\epsilon)\otimes(\p^a(e^A)\Sigma_a\Sigma_9\eta),\label{eqn:connection1}\\
&&\nabla_a(\epsilon\otimes\eta)=\epsilon\otimes(\nabla_8)_a\eta.\label{eqn:connection2}
\eea
These identities can be derived from the local formula
$$\nabla_Q=\p_Q+\frac{1}{4}\omega_{Q}^{RS}\Gamma_R\Gamma_S.$$
Using formulae in Appendix, we have
$$\omega_{\tilde{\alpha}}^{\tilde{\beta}\tilde{\gamma}}=e^{-A}(\omega')_{\alpha}^{\beta\gamma},\quad \omega_{\tilde{\alpha}}^{a\tilde{\beta}}=-\omega_{\tilde{\alpha}}^{\tilde{\beta}a}=-\delta_\alpha^\beta\p^aA,\quad \omega_a^{bc}=(\omega')_a^{bc}$$
and all other components of connection are zero. Therefore
\bea
\nabla_{\tilde{\alpha}}&=&\p_{\tilde{\alpha}}+\frac{1}{4}(\omega_{\tilde{\alpha}}^{\tilde{\beta}\tilde{\gamma}}\Gamma_{\tilde{\beta}}\Gamma_{\tilde{\gamma}}+2\omega_{\tilde{\alpha}}^{a\tilde{\beta}} \Gamma_a\Gamma_{\tilde{\beta}})\nonumber\\ &=&\p_{\tilde{\alpha}}+\frac{1}{4}(e^{-A}(\omega')_{\alpha}^{\beta\gamma}\gamma_\beta\gamma_\gamma\otimes\Sigma_9^2-2\delta_\alpha^\beta\p^a A\,\gamma_\beta\otimes\Sigma_a\Sigma_9)\nonumber\\ &=& e^{-A}(\p_\alpha+\frac{1}{4}(\omega')_{\alpha}^{\beta\gamma}\gamma_\beta\gamma_\gamma)-\frac{1}{2}\gamma_\alpha\otimes(\p^a A\,\Sigma_a\Sigma_9)\nonumber\\ &=& e^{-A}\left((\nabla_3)_\alpha+\frac{1}{2}\gamma_\alpha\otimes\p^a(e^A)\Sigma_a\Sigma_9\right).\nonumber
\eea
Consequently we get (\ref{eqn:connection1}). Similarly (\ref{eqn:connection2}) holds as well.

Moreover, we find, by calculation,
\bea
&&\frac{1}{288}F_{PQRS}(\Gamma_{~~~~~~\alpha}^{PQRS}+8\Gamma^{PQR}\delta_{~\alpha}^S) =\frac{e^{-2A}}{6}\gamma_\alpha\gamma_4\otimes\p^bf\,\Sigma_b,\nonumber\\
&&\frac{1}{288}F_{PQRS}(\Gamma_{~~~~~~a}^{PQRS}+8\Gamma^{PQR}\delta_{~a}^S)= \frac{e^{-3A}}{24}\gamma_4\otimes(\p^b f(\Sigma_b\Sigma_a-\Sigma_a\Sigma_b)-4\p_af)\Sigma_9,\nonumber
\eea
where $\gamma_4=\gamma_1\gamma_2\gamma_3$ is a central element in $\Cl(2,1)$ square to 1.

As for a pinor $\epsilon$ on $M^3$, we have $\gamma_4\epsilon=\pm\epsilon$. Without loss of generality, one may assume $\gamma_4\epsilon=\epsilon$, since this sign corresponds a choice of the pinor bundle ${\cP}_3^\pm$, which gives isomorphic spinor bundle ${\cS}_3$.

With all these preparation, we may compute the curvature tensor $\cF$ of the twisted connection $\cD$. It is straightforward to compute that
$$\cF_{\alpha\beta}(\epsilon\otimes\eta)=(\gamma_{\alpha\beta}\epsilon)\otimes \left(\lambda+\frac{e^{-4A}}{18}\left(|\nabla_8 f|^2-|\nabla_8 e^{3A}|^2-\p^b(e^{3A})\p^cf(\Sigma_b\Sigma_c-\Sigma_c\Sigma_b)\Sigma_9\right)\right)\eta.$$
Now suppose that $(g_3,g_8,A,f)$ is supersymmetric, therefore there exists a spinor $\xi$ such that $\cD\xi=0$. Since ${\cS}_{11}\cong {\cS}_3\otimes {\cP}_8$ and that ${\cS}_3$ is 2-dimensional, we can find a local frame $\epsilon_1$, $\epsilon_2$ of ${\cS}_3$ and write
$$\xi=\epsilon_1\otimes\eta_1+\epsilon_2\otimes\eta_2.$$
In general, $\eta_1$ and $\eta_2$ are combinations of sections of ${\cP}_8$ with function (may have $M^3$ dependence) coefficients. However at any fixed point, we can think of $\eta_1$ and $\eta_2$ as pinors on $M_8$. Since $\cD\xi=0$, we know that
\bea
\cF_{\alpha\beta}(\xi)&=&\gamma_{\alpha\beta}\epsilon_1\otimes\left(\lambda+\frac{e^{-4A}}{18} \left(|\nabla_8 f|^2-|\nabla_8 e^{3A}|^2-\p^b(e^{3A})\p^cf(\Sigma_b\Sigma_c-\Sigma_c\Sigma_b)\Sigma_9\right)\right)\eta_1\nonumber\\
&+&\gamma_{\alpha\beta}\epsilon_2\otimes\left(\lambda+\frac{e^{-4A}}{18}\left(|\nabla_8 f|^2-|\nabla_8 e^{3A}|^2-\p^b(e^{3A})\p^cf(\Sigma_b\Sigma_c-\Sigma_c\Sigma_b)\Sigma_9\right)\right)\eta_2\nonumber\\
&=&0\nonumber
\eea
for any $\alpha,\beta$. In particular, we may choose $\alpha$ and $\beta$ properly such that $\gamma_{\alpha\beta}\epsilon_1$ and $\gamma_{\alpha\beta}\epsilon_2$ are linearly independent, therefore we conclude that
$$\left(\lambda+\frac{e^{-4A}}{18}\left(|\nabla_8 f|^2-|\nabla_8 e^{3A}|^2-\p^b(e^{3A})\p^cf(\Sigma_b\Sigma_c-\Sigma_c\Sigma_b)\Sigma_9\right)\right)\eta_j=0$$
for $j=1,2$.

\smallskip

{\bf Claim}: $\p^b(e^{3A})\p^cf(\Sigma_b\Sigma_c-\Sigma_c\Sigma_b)=0$, or equivalently, there exists a function $h$ such that $\p^b(e^{3A})=h\,\p^bf$ for any index $b$.

\smallskip

We establish the claim.
As $\xi=\epsilon_1\otimes\eta_1+\epsilon_2\otimes\eta_2\neq0$, we may assume that $\eta_1\neq0$ and decompose $\eta_1=\eta_1^+ + \eta_1^-$ as a sum of eigenvectors of $\Sigma_9$. Without loss of generality, we may assume that $\eta_1^+\neq0$, hence, by making use of $\Sigma_b\Sigma_c=2\delta_{bc}-\Sigma_c\Sigma_b$, one obtain
$$\p^b(e^{3A})\p^cf\,\Sigma_c\Sigma_b\,\eta_1^+=\left(\p^b(e^{3A})\p^cf\delta_{bc}-9e^{4A}\lambda+\frac{1}{2}(|\nabla_8 e^{3A}|^2-|\nabla_8 f|^2)\right)\eta_1^+$$
Therefore at any given point we may write
$$\p^b(e^{3A})\p^cf\,\Sigma_c\Sigma_b\,\eta_1^+=\mu\,\eta_1^+$$
for some number $\mu$. By our assumption $f$ is not a constant so we may choose a point such that $\nabla_8 f\neq0$, hence by multiplying $\nabla_8f=\p^a f\Sigma_a$ from left on both sides, we get
$$\left(\p^b(e^{3A})-\frac{\mu\,\p^b f}{|\nabla_8 f|^2}\right)\Sigma_b\,\eta_1^+=0.$$
Consequently
$$\p^b(e^{3A})=\frac{\mu}{|\nabla_8 f|^2}\p^bf$$
for any $b$, the claim is proved, and
\be\label{eqn:relation}
\lambda=\frac{e^{-4A}}{18}(|\nabla_8 f|^2-|\nabla_8 e^{3A}|^2).
\ee

\smallskip

We can compute other components of $\cF$ as well. For example, by making use of the relation $\p^b(e^{3A})\p^cf(\Sigma_b\Sigma_c-\Sigma_c\Sigma_b)=0$, one obtains
\bea
&&(\cF)_{\alpha a}(\epsilon\otimes\eta)\nonumber\\
=&&\gamma_\alpha\epsilon\otimes\left(\frac{e^{-5A}}{18}\p_a f(\p^b(e^{3A})+\p^bf\Sigma_9)\Sigma_b\eta-(\nabla_8)_a\left(\frac{e^{-2A}}{6}(\p^b(e^{3A})\Sigma_9-\p^bf)\right)\Sigma_b\eta\right).\nonumber
\eea
By a similar argument, we see that
\be\label{eqn:comp}
e^{-5A}\p_a f(\p^b(e^{3A})\Sigma_b-\p^bf\Sigma_b\Sigma_9)\eta_j-3(\nabla_8)_a\left(e^{-2A}(\p^b(e^{3A})\Sigma_b\Sigma_9-\p^bf\Sigma_b)\right)\cdot\eta_j=0
\ee
for $j=1,2$ and any $a$. We may assume that $\eta_1^+\neq0$ as before. As we have shown that $\p^b(e^{3A})=h\,\p^bf$ for some function $h$, the above equation can be rewritten as
$$(h-1)(\nabla_8)_a(\p^bf\Sigma_b)\,\eta_1^+=H\,\p^bf\Sigma_b\,\eta_1^+$$
for some smooth function $H$.

\smallskip

{\bf Claim}: $h\equiv 1$.

\smallskip

If the claim is not true, then we can find an open set such that $h-1\neq0$ in that open set. Thus in this open set, we have
$$(\nabla_8)_a(\nabla _8f)=\frac{H}{h-1}\nabla_8 f$$
for any $a$. By pairing with the vector field $e_b$, we get
$$\frac{H}{h-1}f_b=(\nabla_8^2f)_{ab}=\frac{H}{h-1}f_a.$$
As $f$ is not a constant and the frame $\{e_a\}_{a=1}^8$ is arbitrary, the above equation holds only when $H=0$, as $h\neq1$, we get $(\nabla_8^2)f=0$, hence $|\nabla_8 f|^2$ is a nonzero constant. Plugging it back to (\ref{eqn:comp}), one obtain
$$(2h+1)(h-1)\p_aA=h\p_a(h-1).$$
Notice that (\ref{eqn:relation}) now becomes $e^{-4A}(h-1)$ is a constant proportional to $\lambda$. As $h\neq1$, the only possibility is that $A$ is a constant and $h=0$. Plug in (\ref{eqn:comp}) we get $\p^af=0$, contradiction!

So the claim is proved and we conclude that $\nabla_8e^{3A}=\nabla_8 f$. If we work with $\eta_1^-$ instead, then analogously we show that $\nabla_8 f=-\nabla_8 e^{3A}$. As a result, we have shown that supersymmetry implies that
$$df=\pm d(e^{3A}),$$
which further dictates $\lambda=0$ from (\ref{eqn:relation}).

As $\lambda=0$, the Ricci-flatness in dimension 3 implies that $g_3$ is flat, therefore we may choose $\epsilon_1$ and $\epsilon_2$ covariantly constant under $\nabla_3$. In this way, one can show that $\cD\xi=\cD(\epsilon_1\otimes\eta_1+\epsilon_2\otimes\eta_2)=0$ if and only if $\cD(\epsilon_1\otimes\eta_1)=\cD(\epsilon_2\otimes\eta_2)=0$. Therefore we may assume that $\xi=\epsilon\otimes\eta$ is decomposable. Furthermore (\ref{S1}) implies that $\Sigma_9\eta=\pm\eta$, that is, $\eta$ must be a section of one of the spinor bundles ${\cS}_8^\pm$ instead of a random section of the pinor bundle ${\cP}_8={\cS}_8^+\oplus {\cS}_8^-$.

Taking all these into account, we find that $\cD(\epsilon\otimes\eta)=0$ if and only if
\be
(\nabla_8)_a\eta+\frac{1}{8}(4\p_a A+\p^bA(\Sigma_a\Sigma_b-\Sigma_b\Sigma_a))\eta=0\label{eqn:8spinor}
\ee
for any $a$. Consider the conformally changed metric $\bar{g}_8=e^A g_8$, we can identify its Levi-Civita connection $\bar{\nabla}_8$ as\footnote{This formula is well-known, see for example \cite[pp. 16-17]{baum1991}, where we need to change the sign of $\p^bA$ as Baum et al. use the convention $v\cdot v=-q(v)$ for Clifford algebra.}
$$(\bar{\nabla}_8)_a\eta=(\nabla_8)_a\eta+\frac{1}{8}\p^bA(\Sigma_a\Sigma_b-\Sigma_b\Sigma_a)\eta.$$
So (\ref{eqn:8spinor}) can be rewritten as
$$(\bar{\nabla}_8)_a\eta+\frac{1}{2}\p_a A\cdot\eta=0$$
or equivalently
$$\bar{\nabla}_8(e^{A/2}\eta)=0.$$
Thus we have proved that a supersymmetric tuple $(g_3,g_8,A,f)$ implies that $g_3$ is flat, $d f=\pm d(e^{3A})$, and that the conformally changed metric $\bar{g}_8=e^Ag_8$ admits covariantly constant spinors with respect to Levi-Civita connection. The other direction is straightforward.

As a corollary, we have
\begin{corollary}
Suppose $(g_3,g_8,A,f)$ is supersymmetric. Then the conformally changed metric $\bar{g}_8=e^Ag_8$ is Ricci-flat. Moreover, let $N_{11}$ be the number of independent spinors satisfying (\ref{eqn:kspinor}) and let $N_8^\pm$ be the dimension of the space of covariantly constant spinors of the spinor bundle ${\cS}_8^\pm$ associated to the metric $\bar{g}_8$. Then
$$N_{11}=2N_8^+ \textrm{\quad\quad if \quad\quad} df=d(e^{3A})$$
and
$$N_{11}=2N_8^- \textrm{\quad\quad if \quad\quad} df=-d(e^{3A}).$$
\end{corollary}

From Theorem 1 we know that
\be\label{S1}
df=\pm d(e^{3A})
\ee
is a very natural condition, under which we have the following result:

\begin{theorem}\label{theorem2}
Assume (\ref{S1}). Then $(g_3,g_8,A,f)$ is a solution to equations of motion (\ref{eqn:form}), (\ref{eqn:curv}) if and only if

{\rm (a)} $g_3$ is flat,

{\rm (b)} the conformally changed metric $\bar{g}_8=e^Ag_8$ is Ricci-flat,

{\rm (c)} and $A$ satisfies the Laplace equation $\Delta_8A=0$, or equivalently,
\be
\label{S3}
\Delta_{\bar g_8}e^{-3A}=0,
\ee
where $\Delta_8$ and $\Delta_{\bar g_8}$ are the Laplace operators defined respectively by the metrics $g_8$ and $\bar g_8$.
\end{theorem}

\noindent{\it Proof.} Under (\ref{S1}), (\ref{eqn:harmonic}) is equivalent to (\ref{S3}). Under (\ref{S1}) and (\ref{S3}), (\ref{eqn:3curv2}) is equivalent to that $g_3$ is flat. Moreover, (\ref{eqn:8curve}) is equivalent to that $\bar{g}_8$ is Ricci-flat under (\ref{S1}) and (\ref{S3}). Formulae in Appendix are used to derive these equivalences.
\medskip

Summarizing Theorem 1 and Theorem 2, we have proved
\begin{theorem}
\label{theorem3}
The tuple $(g_3,g_8,A,f)$ is a supersymmetric solution to 11D supergravity equation if and only if

{\rm (a)} $g_3$ is flat;

{\rm (b)} $\bar{g}_8:=e^Ag_8$ is a Ricci-flat metric admitting covariantly constant spinors (with respect to Levi-Civita connection);

{\rm (c)} $e^{-3A}$ is a harmonic function on $(M^8,\bar{g}_8)$ with respect to the metric $\bar g_8$;

{\rm (d)} $df=\pm d(e^{3A})$.

\end{theorem}

In \cite{wang1989} McKenzie Wang showed that a simply-connected irreducible Riemannian manifold admits covariantly constant spinors if and only if it has Ricci-flat holonomy, which in dimension 8 must be one of the groups $\mathrm{SU(4)}$, $\mathrm{Sp}(2)$ and $\mathrm{Spin}(7)$. Combining Wang's theorem with Theorem 3, we have the following holonomy classification result. For simplicity, we only state the irreducible and simply-connected case. For more complicated cases, one can consult \cite{moroianu2000} and other references in literature.
\begin{theorem}
Let $(g_3,g_8,A,f)$ be a supersymmetric solution to 11D supergravity on $M^{11}=M^3\times M^8$ with $M^8$ simply-connected and $\bar{g}_8=e^Ag_8$ irreducible. Then one of the following cases must occur:

{\rm (a)} $N_{11}=2$: the metric $\bar{g}_8$ has holonomy group $\mathrm{Spin}(7)$ and $df=d(e^{3A})$;

{\rm (b)} $N_{11}=4$: the metric $\bar{g}_8$ has holonomy group $\mathrm{SU}(4)$ and $df=d(e^{3A})$;

{\rm (c)} $N_{11}=6$: the metric $\bar{g}_8$ has holonomy group $\mathrm{Sp}(2)$ and $df=d(e^{3A})$.

\end{theorem}

We remark that the relation $df=-d(e^{3A})$ may hold in the case $\bar{g}_8$ is reducible. For example, when the solution has maximal number of supersymmetries, i.e., $N_{11}=16$, the Ricci-flat metric $\bar{g}_8$ has to be flat and both cases of $df=\pm d(e^{3A})$ can occur, as in the Duff-Stelle case \cite{duff1991}.

\section{Examples from Ricci-flat manifolds and Green's functions}
\setcounter{equation}{0}

Theorem \ref{theorem2} shows that solutions of $11D$ supergravity can be constructed from a Ricci-flat manifold $\bar M^8$ equipped with a positive harmonic function $e^{-3A}$.
Because the case of constant $A$ would just lead back to the vacuum Einstein equation, we look for $g_8$ being defined on a non-compact manifold, in which case we can just choose $e^{-3A}$ to be a Green's function on a complete manifold. Recall that for a Riemannian manifold $M$, taken to be eight-dimensional for our purposes, a Green's function $G(x,y)$ is a smooth function on $M\times M$ away from the diagonal $x=y$, which is symmetric, harmonic in each variable, positive, and with the following asymptotic
\bea
G(x,y)=d(x,y)^{-6}(1+o(1))
\eea
for $d(x,y)$ small, where $d(x,y)$ is the distance from $x$ to $y$. Green's functions have been shown to exist by Cheng-Li \cite{CL} on compact manifolds, and by Li-Yau\cite{LY} on complete Riemannian manifolds with non-negative Ricci curvature and faster than quadratic volume growth, in the sense that
\bea
{\mathrm{Vol}}(B(p,r))\ge \theta r^{2+\varepsilon}
\eea
for some $\theta>0$ and $\varepsilon>0$. Here $B(p,r)$ the ball of radius $r$ centered at $p\in M$.
Combining these results with Theorem \ref{theorem2}, we obtain the following:

\begin{theorem}\label{theorem5}
Let $(\bar M^8,\bar g_8)$ be a Ricci-flat Riemannian manifold which is either compact or complete with faster than quadratic volume growth. Let $p_1,\cdots,p_n$ be any finite set of points in $\bar M^8$, and $m_1,\cdots,m_n$ a finite set of positive numbers. Set
\bea
G(x)=\sum_{j=1}^n m_j\,G(x,p_j).
\eea
Then the manifold $M^8=\bar M^8\setminus\{p_1,\cdots,p_n\}$, together with the function $A$ defined by $e^{-3A}=G$, gives a solution to the $11D$ supergravity equations. Furthermore the metric $g_8=e^{-A}\bar g_8$ is complete on $M_8$.
\end{theorem}

\bigskip

\noindent{\em Proof:} If $(M^8,\bar{g}_8)$ is {\bf complete with faster than quadratic volume growth}, we have the following estimates for the volume of $B(p,r)$,
\bea
\theta\,r^{2+\varepsilon}
\leq {\mathrm{Vol}}(B(p,r))\leq \omega_8\,r^8
\eea
where the upper bound is a consequence of the Bishop-Gromov volume comparison theorem. By a result of Li and Yau \cite{LY},
there exists a Green's function $G(x,y)$ on $(\bar M_8,\bar g_8)$ satisfying
\bea \label{eqn:LY}C^{-1}\int_{d(x,y)^2}^\infty \frac{1}{{\mathrm{Vol}}(B(x,\sqrt{t})  )}dt \le G(x,y) \le  C\int_{d(x,y)^2}^\infty  \frac{1}{{\mathrm{Vol}}(B(x,\sqrt{t})  )}dt
\eea  for some uniform constant $C>1$. It follows that
\bea \label{eqn:Green}C^{-1} \frac{1}{d(x,y)^6}\le G(x,y)\le C \theta^{-1} \frac{1}{d(x,y)^\varepsilon},\quad {\mathrm{for\,\, any}}\,\, x,y\in \bar M_8 .
\eea

Recall that $e^{-A} = G^{1/3}$ and $g_8= G^{1/3} \bar g_8$. The metric $g_8$ is a well-defined Riemannian metric on $M^8$, the complement of the singular points $p_1,\ldots, p_n$ in $\bar M^8$. Near each singular point $p_i$ $$g_8\sim \frac{1}{d_{\bar g_8}(p_i,x)^2} \bar g_8.$$ Therefore  $d_{g_8}(x,p_i) = +\infty$ for any $x\in M^8$ and $(M^8,g_8)$ is complete near each singular point $p_i$. Moreover, the metric $g_8$ is asymptotically to the standard product metric on $\R \times S^7$ as $d_{\bar g_8}(p_i, x)\to 0$.

Similarly, the completeness of $(M^8,g_8)$ near infinity is a consequence of the estimates (\ref{eqn:Green}) for the Green's function. So $(M^8,g_8)$ is a complete Riemannian manifold.

\bigskip

If $\bar M^8$ is {\bf compact}, we may take $e^{-3A} = G$. By similar estimates as above, we see that $g_8 = e^{-A} \bar g_8$ is complete near each singular point $p_i$, and in this case, $(M^8, g_8)$ is also complete.

\bigskip

We survey some known examples of compact or noncompact Ricci-flat $8$-dimensional manifolds. By Theorems \ref{theorem2} and \ref{theorem3}, we have complete solutions to the field equations (\ref{eqn:harmonic}), (\ref{eqn:8curve}) and (\ref{eqn:3curv}) for such examples. They are all supersymmetric, except possibly for some examples constructed from Riemann surfaces in section 3.3.

\subsection{Compact examples}
The following examples are included in Joyce's book \cite{Jo}.

\medskip

\noindent{\bf 1.} Compact real $8$-manifolds with holonomy ${\mathrm{Spin}}(7)$. Let $T^8$ be a torus equipped with a flat ${\mathrm{Spin}(7)}$-structure $(\Omega_0,g_0)$, and a finite group $\Gamma$ of automorphisms of $T^8$ preserving $(\Omega_0,g_0)$. Then $T^8/\Gamma$ is an orbifold with flat Spin$(7)$-structure $(\Omega_0, g_0)$. Let $\bar M^8$ be a suitable resolution of $T^8/\Gamma$ such that $\bar M^8$ is simply connected. Then $\bar M^8$ admits a torsion free Spin$(7)$-structure $(\bar \Omega, \bar g)$ with Hol$(\bar g) = {\mathrm{Spin}(7)}$. Thus $\ric(\bar g) = 0$. More examples of compact $\textrm{Spin}(7)$ manifolds can be found in \cite{Cl, Ta}

\medskip

\noindent{\bf 2. } Compact  real $8$-manifolds with holonomy ${\mathrm{Sp}}(2)$. Here we briefly present two examples of Beauville \cite{B}. We start with a compact complex surface $X$. Let $X^{(m)}$ be the $m^{{\mathrm{th}}}$-symmetric product of $X$ which is a complex orbifold with complex dimension $2m$. Take $X^{[m]}$ to be the {\em Hilbert scheme} of $0$-dimensional subspaces $(Z,{\mathcal O}_Z) $ of $X$ of length dim$_{{\mathbf C}} {\mathcal O}_Z = m$. Then $X^{[m]}$ is a compact complex manifold with dim$_{{\mathbf C}} X^{[m]} = 2m$, and the natural projection $\pi: X^{[m]}\to X^{(m)}$ is a crepant resolution. (1) If $X$ is a $K3$-surface, then $X^{[2]}$ admits a $61$-dimensional family of metrics $\bar g$ with holonomy ${\mathrm{Sp}}(2)$. (2) If $X$ is a compact complex torus $T^4$ which can be regarded as an abelian Lie group. So there is a natural map $\sigma: X^{(3)}\to  X$ given by the summing the $3$ points. Let $K^2(X)$ be the kernel of the map $\sigma\circ \pi: X^{[3]}\to X$, then $K^2(X)$ is a a complex $4$-dimensional manifold admitting a $13$-dimensional family of metrics $\bar g$ with holonomy ${\mathrm{Sp}}(2)$.

\medskip

\noindent{\bf 3. } Compact real $8$-manifolds with holonomy ${\mathrm{SU}}(4)$. These are complex $4$-dimensional Calabi-Yau manifolds with trivial first Chern class. By Yau's theorem \cite{Y} there is a unique Ricci-flat K\"ahler metric in each K\"ahler class. Examples of Calabi-Yau $4$-manifolds include smooth hypersurfaces in ${\mathbf{CP}}^5$ with degree $6$.

\subsection{Non-compact examples, complete with maximum volume growth}

In this section we will present some examples of complete Ricci-flat metrics on noncompact $8$-manifolds with maximal volume growth.

\medskip

\noindent{\bf 1. } {\em Complete K\"ahler-Ricci-flat metrics on ${\mathbf C}^4$.}  Recently, Sz\'ekelyhidi \cite{Sz}, Conlon-Rochon \cite{CR} and Li \cite{Li} constructed nontrivial Ricci-flat K\"ahler metrics on ${\mathbf C}^4$ with maximal volume growth. The desired Ricci-flat metrics $\bar g$ are  perturbations of (singular) Ricci-flat metrics on some metric cone to which  $({\mathbf C}^4,\bar g)$ is asymptotic. More precisely, let $f$ be a polynomial on ${\mathbf C}^4$ and $M_1 \subset {\mathbf C}^5$ be the graph of $-f$ defined by $z + f(x_1,\ldots, x_4) = 0$. So $M$ is biholomorphic to ${\mathbf C}^4$. Assume the cone $f^{-1}(0)\subset {\mathbf C}^4$ has isolated singularity and the cone $M_0 = {\mathbf C}\times f^{-1}(0)$ admits (singular) Ricci-flat cone metric $g_0$, then $g_0$ can be perturbed to a complete K\"ahler-Ricci-flat metric $g_1$ on $M_1$, with tangent cone at infinity isometric to $(M_0,g_0)$, in particular, $(M_1,g_1)$ has maximal volume growth and is non-trivial for generic choice of $f$. For example, if $f(x) = x_1^2 + x_2^2 + x_3^2 + x_4^2$, i.e. $f^{-1}(0)$ is the $A_1$-singularity, $M_0$ admits a Ricci-flat cone metric given $g_{{\mathrm{St}}}$ by the product of Stenzel metric and standard metric on ${\mathbf C}$. So ${\mathbf C}^4$ admits a nontrivial complete Ricci-flat K\"ahler metric which is asymptotic to $(M_0,g_{{\mathrm{St}}})$.

\medskip

\noindent{\bf 2. } {\em Complete K\"ahler-Ricci-flat metrics on quasi-projective manifolds}. This class of Ricci-flat metrics was first constructed by Tian-Yau \cite{TY} and later refined by Conlon-Hein \cite{CH}. If $\bar M$ is a compact  K\"ahler orbifold with dim$_{{\mathbf C}} (\bar M) = 4$ and codim$_{{\mathbf C}}({\mathrm{Sing}}( \bar M)) \ge 2$. Let $D$ be a {\em neat} and {\em almost ample} sub-orbifold divisor in $\bar M$ such that Sing$(\bar M)\subset D$ and $D \in |- \beta K_{\bar M}| $ for some $\beta\in (0,1)$. If $D$ admits a KE metric with positive scalar curvature, then $M = \bar M\backslash D$ admits a complete Ricci-flat metric with maximal volume growth.

\medskip

\noindent{\bf 3. } {\em Asymptotical conic Ricci-flat manifolds.} Let $M$ be a complex noncompact manifold with a nontrivial holomorphic $4$-form $\Omega$ and $(M,\Omega)$ is asymptotic to a Calabi-Yau cone $(C,g_0,\Omega_0)$ with some positive rate. Then it is shown by Conlon-Hein \cite{CH1} that there exists a unique Ricci-flat metric $\bar g$ in each K\"ahler class on $M$ with suitable asymptotic condition, such that $\bar g$ is asymptotic to the cone metric $g_0$ with some positive rate. Therefore $(M,\bar g)$ has maximal volume growth. Examples of this type include the ALE K\"ahler complex dimension $4$-manifolds studied by Joyce \cite{Jo}.



\subsection{The case of Riemann surfaces}
In this section, we look for solutions on $M^8 =M^6\times M^2 $ with Riemannian metrics of the form
\bea \nonumber
g_8 = e^{2B}g_6 + g_2,\quad B\in C^\infty(M^2).
\eea
We also assume the $1$-form $df$ and the function $A$ depend only on $M^2$. We rewrite the equations (\ref{eqn:8curve}) and (\ref{eqn:3curv2}) with $\lambda = 0$ as (we use $i,\, j,\ldots$ to denote the indices on $M^2$ and $\mu,\nu,\ldots$ those on $M^6$)
\bea\nonumber
\ric(g_2)_{ij} - 6 (\nabla^2_{g_2} B)_{ij} - 6 B_i B_j - 3 (\nabla^2_{g_2} A)_{ij} - 3 A_i A_j + \frac{e^{-6A}}{2} f_i f_j - \frac{e^{-6A}}{6}|\nabla f|^2_{g_2} (g_2)_{ij} = 0,
\eea
\bea\label{eqn:ric6}
\ric(g_6)_{\mu\nu} - (\Delta_{g_2} B + 6|\nabla B|^2  ) e^{2B} g_{6,\mu\nu} - 3 g_2( \nabla A, \nabla B) e^{2B} g_{6,\mu\nu} - \frac{e^{-6A}}{6} |\nabla f|^2_{g_2} e^{2B} g_{6,\mu\nu} = 0,
\eea
and
\bea\nonumber
\Delta_{g_2} A + 6 g_2(\nabla A,\nabla B) + 3 |\nabla A|^2_{g_2} - \frac{e^{-6A}}{3} |\nabla f|^2_{g_2} = 0.
\eea
Since $A$, $B$ and $f$ depend only on $M_2$, equation (\ref{eqn:ric6}) implies $g_6$ is an Einstein metric, $\ric(g_6) = \hat \lambda g_6$ for some $\hat\lambda \in {\R}$, and (\ref{eqn:ric6}) becomes
\bea\nonumber
\Delta_{g_2} B + 6 |\nabla B|^2 + 3 g_2(\nabla A,\nabla B) + \frac{e^{-6A}}{6} |\nabla f|^2 = \hat\lambda e^{-2B}.
\eea
With the supersymmetry assumption (\ref{S1}), $df = \pm d(e^{3A})$, the equations above are reduced to
\bea\label{eqn:ric2}
 \ric(g_2)_{ij} - 6\nabla^2_{ij} B - 6 B_i B_j - 3 \nabla^2_{ij} A + \frac{3}{2} A_i A_j - \frac{3}{2}|\nabla A|^2 g_{2,ij} = 0,
\eea
\bea\label{eqn:B1}
\Delta_{g_2} B + 6 |\nabla B|_{g_2}^2 + 3 g_2(\nabla A, \nabla B) + \frac{3}{2} |\nabla A|_{g_2}^2 = \hat\lambda e^{-2B},
\eea
and
\bea\label{eqn:A1}
\Delta_{g_2} A + 6 g_2(\nabla A,\nabla B) = 0.
\eea
If furthermore we assume $(M^6,g_6)$ is Ricci-flat, i.e. $\hat \lambda = 0$, then one can check equations (\ref{eqn:B1}) and (\ref{eqn:A1}) are equivalent to
\bea\label{eqn:F1}
\Delta_{g_2} K + 6 |\nabla K|_{g_2}^2 = 0,\quad {\mathrm{or}} \quad \Delta_{g_2} e^{6 K} = 0,
\eea
and
\bea\label{eqn:A2}
\Delta_{g_2} e^{-3A} + 6 g_2(\nabla K,\nabla e^{-3 A}) = 0,
\eea
where we denote $K = B  + \frac{A}{2}$. And equation (\ref{eqn:ric2}) becomes
\bea \label{eqn:ric 3}
\ric(g_2)_{ij} - 6( \nabla^2_{g_2}K)_{ij} - 6 K_i K_j + 3 K_i A_j + 3 K_j A_i - \frac{3}{2}|\nabla A|_{g_2}^2 g_{2,ij} = 0.
\eea
With equations (\ref{eqn:F1}), (\ref{eqn:A2}) and (\ref{eqn:ric 3}) at hand, we discuss some explicit solutions. Recall that we fix a Ricci-flat manifold $(M^6, g_6)$.

\bigskip

\noindent{\bf 1. } When $K=\;{\mathrm{const}}$, we take $e^{-3A}$ to be a positive (possibly singular) harmonic function on $M^2$ (note that the choice of harmonic functions on $M^2$ is independent of the metric $g_2$). The equation (\ref{eqn:ric 3}) implies that $$\ric(g_2)_{ij} - \frac{3}{2} |\nabla A|_{g_2}^2 g_{2,ij} = \ric(g_2)_{ij} - \frac{1}{2}\Delta_{g_2} A  g_{2,ij} = 0.$$
If we define $\bar g_2 = e^{A} g_2$, the equation above says that $\ric(\bar g_2) = 0$ on the set of $M^2$ where $e^{3A}$ is smooth.  As remarked above, $\Delta_{\bar g_2}e^{-3 A} = 0$. This implies the product metric $\bar g_8 = e^{A} g_8  = e^{2 K} g_6 + e^{A } g_2$ is Ricci-flat, which is just a special case of Theorem \ref{theorem2}. This solution can be made supersymmetric by taking $M^6$ to be Calabi-Yau.

\medskip

By the uniformization theorem, complete flat manifolds $(M^2,\bar g_2)$ are either the Euclidean plane ${\mathbf C}$, cylinder $S^1\times {\R}$ or the compact torus $T^2$. Since such manifolds cannot admit non-constant {\bf smooth positive} harmonic functions, we cannot expect $e^{-3 A}$ to be smooth. Since ${\mathbf C}$ is parabolic, i.e. it admits no positive Green function, for such examples, we can take $e^{-3A}$ to be constant, which gives rise to trivial solutions.

\medskip

To get nontrivial solutions, we choose $M^2$ to be an open Riemann surface with boundary $\partial M^2\neq \emptyset$. For example we may take $M^2$ to be the unit disk $D\subset{\mathbf C}$ or the punctured unit disk $D^*\subset{\mathbf C}\backslash\{0\}$. We pick $\bar g_2 = g_{{\mathbf C}}$, the Euclidean metric on ${\mathbf C}$, and $e^{-3A} = {\mathrm{Re}}(\phi) -  \mu \log |z|^2$ for any holomorphic function $\phi\in {\mathcal O}(D)$ with positive real part ${\mathrm{Re}}(\phi)>0$ and any $\mu\ge 0$. Then $g_2 = e^{-A} g_{{\mathbf C}} $ defines a solution to (\ref{eqn:ric 3}). However, the metric $g_2$ is incomplete on $D$ or $D^*$. In sum, the tuple $(g_{11}, F, A)$ given by
$$ g_{11} = e^{2A} g_3 + e^{2K - A} g_6 + e^{-A} g_{{\mathbf C}}, \quad F = \pm dvol_3\wedge d(e^{3A})$$
$$A = -\frac 1 3 \log \big( {\mathrm{Re}}(\phi) - \mu \log|z|^2  \big),\quad \forall\; \phi\in {\mathcal O}(D), \, \mu\ge 0,$$
on $M^{11} = M^3\times M^6\times M^2$ with $(M^3,g_3)$ and $(M^6,g_6)$ being Ricci-flat and $M^2 = D$ or $D^*$, satisfy the equations of motion (\ref{eqn:harmonic}), (\ref{eqn:8curve}) and (\ref{eqn:3curv}). Replacing $g_6$ by $e^{2K}g_6$, we can also express $g_{11}$ as
\bea
g_{11}=\big( {\mathrm{Re}}(\phi) - \mu \log|z|^2  \big)^{-{2\over 3}}g_3
+
\big( {\mathrm{Re}}(\phi) - \mu \log|z|^2  \big)^{1\over 3}(g_6+g_{\mathbf C}).
\eea

\bigskip

\noindent{\bf 2. } When $K \neq {\mathrm{const}}$, from (\ref{eqn:F1}), we know $e^{6K}$ is a positive harmonic function. Again we take $M^2 = D$ or $D^*$. Let $z = x_1+ ix_2\in{\mathbf C}$ be the standard coordinate on $D$ or $D^*$. Observe from (\ref{eqn:A2}) that if $e^{-3A}$ is proportional to $K$, (\ref{eqn:A2}) is also satisfied. By adding a positive constant to $K$ if necessary we may assume $K>0$ and $e^{-3A} = K$, i.e. $A = -\frac 1 3 \log K$. With this choice of $(K,A)$, it suffices to find metrics $g_2$ to the equation (\ref{eqn:ric 3}), which can be rewritten as
\bea \label{eqn:ric new}
\ric(g_2)_{ij} - 6( \nabla^2_{g_2}K)_{ij}  - 6 K_i K_j  - 2 \frac{K_i K_j}{K} - \frac{1}{6}\frac{|\nabla K|^2_{g_2}}{K^2} g_{2,ij} = 0.
\eea
This is a system of second order partial differential equations in $g_2$, and we do not expect to find general solutions to this equation for general choice of $e^{6K}$. So we now focus on some special cases depending on the positive harmonic function $e^{6K}$.

\medskip

 $\bullet$ If  $e^{6K}$ is linear in $x_1$ and $x_2$, e.g. $e^{6K} = x_1 + 10$.  We look for the metrics conformal to the Euclidean one, i.e. $g_2 = e^{2\varphi} g_{{\mathbf C}}$ for some $\varphi\in C^\infty(M^2)$, equation (\ref{eqn:ric new}) becomes
\bea\label{eqn:new ric}
&&-\Delta_{g_{\mathbf C}} \varphi\, \delta_{ij}  - 6\,\partial^2_{ij} K + 6\, (\varphi_i K_j + \varphi_j K_i) - 6 \langle \nabla K,\nabla \varphi \rangle_{g_{{\mathbf C}}} \delta_{ij}\nonumber\\
&&\quad  - 6 K_i K_j - 2 \frac{K_i K_j}{K} - \frac{1}{6} \frac{|\nabla K|_{g_{{\mathbf C}}}^2}{K^2} \delta_{ij} = 0 .
\eea
By straightforward calculations, we can check that
\bea\nonumber\varphi = -\frac{5}{2} K + \frac 1 6 \log K + C,\quad {\mathrm{for \, any\, constant\,}} C\in{\R},
\eea satisfies equation (\ref{eqn:new ric}). Hence $g_2 = e^{-5K} K^{1/3} g_{{\mathbf C}}$ defines a solution to (\ref{eqn:ric new}). Therefore the tuple $(g_{11}, F, K)$ given by
$$ g_{11} = K^{-2/3} g_3 + K^{1/3} e^{2K} g_6 + K^{1/3} e^{-5 K} g_{{\mathbf C}},\quad F = \pm dvol_3 \wedge dK$$
satsifies the equations of motion (\ref{eqn:harmonic}), (\ref{eqn:8curve}) and (\ref{eqn:3curv}) on $M^{11} = M^3\times M^6 \times M^2$ with $M^2 = D$ or $D^*$. For example, we may take $K = \frac 16 \log (10 + x_1)$, then
$$g_{11} = \frac{6^{2/3}}{\big(\log(10+ x_1)\big)^{2/3}} g_3 + \frac{ \big((10+x_1)\log(10+x_1)\big)^{1/3}}{6^{1/3}} g_6 + \frac{\big(\log (10+x_1) \big)^{1/3}   }{6^{1/3} (10+x_1)^{5/6}} g_{{\mathbf C}} $$
and
$$F = \frac{\pm 1}{6(10+x_1)} dvol_3 \wedge d x_1$$
define an explicit solution, where we recall that $x_1$ is one of the coordinates on $D$ or $D^*$.


\medskip

$\bullet$ If $e^{6K}$ is radial symmetric, i.e. it depends only on $r = |z|$. For example, we can take $e^{6K} = - c \log r^2 + 1 $ for any $c>0$. Again $A = -\frac 1 3 \log K$ and we try to find metrics $g_2$ of the form $g_2 = e^{2\varphi} g_{{\mathbf C}}$ for some $\varphi\in C^\infty(D^*)$. One can check that $$\varphi = \frac 1 6 \log K - \log r - \frac 5 2 K + C,\quad {\mathrm{for\, any\, constant\,}}C\in{\R}$$ satisfies the equation (\ref{eqn:new ric}). Therefore $g_2 =r^{-2} e^{-5 K + \frac 1 3 \log K } g_{{\mathbf C}}$ satisfies the equation (\ref{eqn:ric 3}) and correspondingly, the tuple $(g_{11}, F)$ given by
$$g_{11} = K^{-2/3} g_3 + K^{1/3}e^{2K} g_6 + r^{-2} K^{1/3} e^{-5 K} g_{{\mathbf C}}, \quad F = \pm dvol_3\wedge dK $$
defines a solution to the equation of motion  (\ref{eqn:harmonic}), (\ref{eqn:8curve}) and (\ref{eqn:3curv})  on $M^{11} = M^3 \times M^6\times D^*$. In particular, if we choose $e^{6K} = - \log r^2+1$ on $D^*$, then $(g_{11}, F)$ are given by
\bea\nonumber
g_{11}= && \frac{6^{2/3}}{\big(\log( 1 - \log r^2 )\big)^{2/3}} g_3 + \frac{\big(\log (1- \log r^2)\big)^{1/3}}{6^{1/3}} (1-\log r^2)^{1/3} g_6\\
&& + r^{-2} \frac{\bigl(\log (1- \log r^2)\bigr)^{1/3}}{6^{1/3}} (1-\log r^2)^{-5/6} g_{{\mathbf C}}\nonumber
\eea
and $$F = \frac{\pm 1}{3r (1-2\log r)} dvol_3\wedge dr.$$




\section{The Duff-Stelle Ansatz}
\setcounter{equation}{0}

In their seminal paper \cite{duff1991}, Duff and Stelle discovered the (multi-)membrane solution to $11D$ supergravity by making following assumptions on the tuple $(g_3,g_8,A,f)$:

\smallskip

(a) $g_3$ is flat;

(b) The Killing spinor $\xi$ is a pure tensor product of a covariantly constant spinor $\epsilon$ on $M^3$ and a pinor $\eta$ on $M^8$;

(c) $M^8$ is a radially symmetric open domain in $\R^8$,
the metric $(g_8)_{ij}=e^{2B}\delta_{ij}$ is conformally flat, where $B$ is a smooth function on $M^8$ and all the functions $A$, $B$ and $f$ depend only on the radial variable $r$.

\medskip

From the analysis in Section 2 we know that assumptions (a) and (b) above are necessary for supersymmetric solutions. In this section, we first re-derive the Duff-Stelle solution using the framework of Section 3. Then we show that, by keeping assumption (c) above only, we can construct a 5-parameter family of solutions to equations of motion, extending Duff-Stelle's work. Due to the classification result (Theorem 3 in Section 2), the only supersymmetric solution in this family is the Duff-Stelle solution.

\subsection{Derivation of the Duff-Stelle membrane solution}

Take $(\bar{M}^8,\bar{g}_8)$ to be $\R^8$ equipped with Euclidean metric in Theorem 5 and let $M^8$ be the punctured Euclidean space $\R^8\setminus\{0\}$. It is well-known that a Green's function on $\R^8$ with source at origin is given by
$$G(x)=\frac{1}{r^6},$$
where $r$ is the Euclidean distance to origin. By taking
$$e^{-3A}=G(x)+M$$
for any nonnegative constant $M$, we get Duff-Stelle's membrane solution described in \cite{duff1991}. If we take $e^{-3A}$ to be a positive linear combination of Green's functions at different points and a positive constant, then we recover the multi-membrane solution.

\subsection{The $5$-parameter family of radially symmetric solutions}

In this subsection, we solve for all solutions $(g_3,g_8,A,f)$ to equations of motion under Assumption (c). As $A$, $B$ and $f$ depend only $r$, (\ref{eqn:form}) and (\ref{eqn:curv}) are reduced to an ODE system:
\bea
&&f''+f'\left(\frac 7 r +6B'-3 A'\right)=0\label{eqn:ds1},\\
&&3e^{2A}\left(A''+A'\left(\frac 7 r +6B'\right)+3(A')^2\right)-e^{-4A}(f')^2=3\lambda e^{2B},\label{eqn:ds2}\\
&&-2B''+\frac{2B'}{r}+2(B')^2-A''+\frac{A'}{r}+2A'B'-(A')^2=-\frac{e^{-6A}}{6}(f')^2,\label{eqn:ds3}\\
&&B''+\frac{13B'}{r}+6(B')^2+\frac{3A'}{r}+3A'B'=-\frac{e^{-6A}}{6}(f')^2.\label{eqn:ds4}
\eea
The goal is to solve this complicated nonlinear ODE system.

First notice that (\ref{eqn:ds1}) can be integrated to
$$f'=Mr^{-7}e^{3A-6B}$$
for some constant $M$. Plug it into other equations, we get
\bea
&&\frac{M^2}{3r^{14}}e^{-12B}=A''+A'\left(\frac 7 r +6B'\right)+3(A')^2-\lambda e^{2B-2A},\nonumber\\
&&\frac{M^2}{3r^{14}}e^{-12B}=4B''-\frac{4B'}{r}-4(B')^2+2A''-\frac{2A'}{r}-4A'B'+2(A')^2,\nonumber\\
&&\frac{M^2}{3r^{14}}e^{-12B}=-2B''-\frac{26B'}{r}-12(B')^2-\frac{6A'}{r}-6A'B'.\nonumber
\eea

Let $L=L(r)$ be a function such that
$$e^B=e^L\cdot r^{-1},$$
hence
$$B'+\frac 1 r=L'.$$
The original system can be rewritten as

\bea
&&\frac{M^2}{3r^2}e^{-12L}=A''+A'\left(\frac{1}{r}+6L'\right)+3(A')^2-\frac{\lambda}{r^2}e^{2L-2A}, \label{eqn:ddds1}\\
&&\frac{M^2}{3r^2}e^{-12L}=4L''+2A''+\frac{4}{r}L'+\frac{2}{r}A'-4(L')^2-4A'L'+2(A')^2+\frac{4}{r^2}, \label{eqn:ddds2}\\
&&\frac{M^2}{3r^2}e^{-12L}=-2L''-\frac{2}{r}L'-12(L')^2-6A'L'+\frac{12}{r^2}. \label{eqn:ddds3}
\eea

The above ODE system (\ref{eqn:ddds1})-(\ref{eqn:ddds3}) has only two unknown functions $A$ and $L$. Therefore this system is overdetermined and a priori it may be inconsistent itself. Surprisingly, we have the following theorem.

\begin{theorem}
The ODE system (\ref{eqn:ddds1})-(\ref{eqn:ddds3}) is consistent. In fact, it is equivalent to a single 3rd order nonlinear ODE
\be
\frac{d^3v}{d t^3}+7\frac{d^2v}{d t^2}v+14\left(\frac{d v}{d t}\right)^2+2\frac{d v}{d t}(17v^2-60) +12(v^2-4)(v^2-6)=0.\label{eqn:3order}
\ee
As a consequence, we get a 5-parameter family of solutions to the equations of motion (\ref{eqn:form}) and (\ref{eqn:curv}) of $11D$ supergravity.
\end{theorem}

\medskip
\noindent{\it Proof}. Write $u=A'+2L'$ and $T=L-A$, we get
\bea
&&u'+\frac{u}{r}+3u^2-\frac{12}{r^2}=\frac{\lambda}{r^2}e^{2T},\label{eqn:a1}\\
&&4u'+T''+\frac{4}{r}u+\frac{1}{r}T'+2u^2+uT'+2(T')^2=\frac{12}{r^2}\label{eqn:a2}
\eea
by eliminating the left hand side of (\ref{eqn:ddds1})-(\ref{eqn:ddds3}). Notice that the second equation can be rearranged as
$$(e^{2T})''+\left(\frac{1}{r}+u\right)(e^{2T})'+4\left(2u'+\frac{2}{r}u+u^2-\frac{6}{r^2}\right)e^{2T}=0.$$
Introduce $v=ur$ and eliminate $e^{2T}$, we get
$$v'''r^3+v''r^2(7v+3)+14(v'r)^2+v'r(34v^2+7v-119)+12(v^2-4)(v^2-6)=0.$$
Let $r=e^t$, then we get the desired 3rd order ODE (\ref{eqn:3order}).

Assume we have a solution $v$ of (\ref{eqn:3order}), then we know $u$ and we can solve for $e^{2T}$ from (\ref{eqn:a1}), hence also $A'$ and $L'$. Therefore $A$ and $L$ are determined up to an additive constant.

To check the consistency of the ODE system, one only needs to verify that the functions $A$ and $L$ we get above satisfy any of the equations in the original ODE system.

We introduce
$$X=\left(\frac{d v}{d t}+3v^2-12\right)^2\left(2\frac{d v}{d t}-v^2+4\right) +\frac{1}{3}\left(\frac{d^2v}{d t^2}+5\frac{d v}{d t}v-3v^3+12v\right)^2.$$
By a lengthy calculation, we find that the consistency condition is that $X$ satisfies
$$\frac{d X}{d t}+4vX=0,$$
which turns out to be a consequence of (\ref{eqn:3order}).

Therefore solving the equations of motion (\ref{eqn:form}) and (\ref{eqn:curv}) under our ansatz is equivalent to solving the 3rd order ODE (\ref{eqn:3order}). As there are 3 parameters for $v$, one additive parameter to determine $A$ and $L$, and an extra parameter $\lambda$, we get a 5-parameter family of solutions to the 11D supergravity.

In general we do not know how to write down all the solutions to (\ref{eqn:3order}), however, there are some explicit special solutions we can find.
\medskip

\noindent{\bf 1. }For $v$ satisfying $$\frac{d v}{d t}=-(v^2-4),$$ (\ref{eqn:3order}) is automatically satisfied. This corresponds to the case $L'=0$, or equivalently $B'=-1/r$. So $L$ is a constant and (\ref{eqn:ddds3}) implies that $M=\pm6e^{6L}$. Furthermore, (\ref{eqn:ddds1}) and (\ref{eqn:ddds2}) become
\bea
&&\frac{12}{r^2}=A''+\frac{A'}{r}+3(A')^2-\frac{\lambda}{r^2}e^{2L-2A},\nonumber\\
&&\frac{4}{r^2}=A''+\frac{A'}{r}+(A')^2.\nonumber
\eea
The general solution is given by
$$e^{2A}=\frac{\lambda e^{2L}}{32C}\frac{(1-Cr^4)^2}{r^4},$$
where $C$ is a constant with the convention that
$$e^{2A}=C_1r^{-4}$$
for $C=0$, in which case $\lambda=0$, and
$$e^{2A}=C_1r^4$$
for $C=\infty$, in which case $\lambda=0$. We can also write down the explicit expressions of $B$ and $f'$. It turns out that these solutions are isometric to either Freund-Rubin solutions or Ricci-flat solutions.
\medskip

\noindent{\bf 2. } For $v$ satisfying $$\frac{d v}{d t}=-(v^2-6),$$ (\ref{eqn:3order}) is automatically satisfied. The corresponding solutions are given by
\bea
&&e^{2A}=\frac{\lambda C_2}{6}\frac{(r^{2\sqrt{6}})^{-1/3}(1-Cr^{2\sqrt{6}})^2}{(1+6Cr^{2\sqrt{6}}+C^2r^{4\sqrt{6}})^{2/3}},\nonumber\\
&&e^{2B}=\frac{C_2}{r^2}\left(\frac{1+6Cr^{2\sqrt{6}}+C^2r^{4\sqrt{6}}}{r^{2\sqrt{6}}}\right)^{1/3},\nonumber\\
&&f'=\pm\frac{(32C^2\lambda^3C_2^3)^{1/2}r^{\sqrt{6}-1} (1-Cr^{2\sqrt{6}})^3}{3(1+6Cr^{2\sqrt{6}}+C^2r^{4\sqrt{6}})^2}.\nonumber
\eea
Here we should use the convention that
\bea
e^{2A}=\frac{\lambda C_2}{6}r^{-2\sqrt{6}/3},\qquad
e^{2B}=C_2r^{-2-2\sqrt{6}/3},\qquad
f'=0,\nonumber
\eea
when $C=0$ and
\bea
e^{2A}=C_1r^{2\sqrt{6}/3},\qquad
e^{2B}=\frac{6C_1}{\lambda}r^{2\sqrt{6}/3-2},\qquad
f'=0,\nonumber
\eea
when $C=\infty$. Here $C_1,C_2,C$ are constants (with relation $C_1=\lambda C_2C^{2/3}/6$). So we get a 3-parameter family of explicit solutions to 11D supergravity with $\lambda$ not necessarily zero. The corresponding metrics $g_8$ on $M^8$ are incomplete.
\medskip

\noindent{\bf 3. } For $v$ satisfying $$\frac{d v}{d t}=-3(v^2-4),$$ (\ref{eqn:3order}) is also automatically satisfied. This corresponds to the case $\lambda=0$ considered in Duff-Stelle \cite{duff1991}. As we are working with field equations only, we get more general solutions compared to Duff-Stelle's result.

What is more interesting is that for the $\lambda=0$ case, the ODE system (\ref{eqn:ds1})-(\ref{eqn:ds4}) can be solved explicitly and completely as follows.

We eliminate $(f')^2$ from (\ref{eqn:ds2}) and (\ref{eqn:ds4}) to get
$$(A''+2B'')+\frac{13}{r}(A'+2B')+3(A'+2B')^2=0.$$
Write $u=A'+2B'$ (this $u$ is slightly different from the previous $u$), so we have
\be
u'+\frac{13u}{r}+3u^2=0.\label{eqn:u}
\ee

\noindent{\bf (a). } If we take $u=0$ as in Duff-Stelle \cite{duff1991}, then (\ref{eqn:ds3}) reduces to
  $$[(e^{3A})']^2=(f')^2,$$
  so
  $$df=\pm d(e^{3A}),$$
  therefore (\ref{eqn:ds1}), (\ref{eqn:ds2}) and (\ref{eqn:ds4}) all reduce to
  \be
  B''+\frac{7B'}{r}+6(B')^2=0,\label{eqn:b}
  \ee
  which implies that the scalar curvature of the metric $g_8$ is nonnegative.

  If $B'=0$, we get the trivial solution with $F$=0 and $M^{11}$ Ricci-flat.
  Another solution is given by
  $$B'=\frac{-1}{r},$$
  hence
  \bea
  e^{2A}=C_1r^4,\qquad
  e^{2B}=\frac{C_2}{r^2},\qquad
  f'=\pm6C_1^{3/2}r^5.\nonumber
  \eea
  We see that $g_8$ is a complete conformally flat metric on $\R^8\setminus\{0\}$ which is isometric to $C_2(\R\times S^7)$. The eleven dimensional manifold $M^{11}$ is isometric to
  $$(M^3\times\R)\times S^7(\sqrt{C_2}),$$
  where the metric on $M^3\times\R$ is given by
  $$g_4=C_1e^{4x/\sqrt{C_2}}g_3+(dx)^2,$$
  where $x$ is the coordinate on $\R$ and $g_3$ is a flat Lorentzian metric on $M^3$. The metric $g_4$ is Einstein satisfying
  $$(\ric_4)_{ij}=-\frac{12}{C_2}(g_4)_{ij}.$$
  In this case one can also check that
  $$F=\pm\frac{6}{\sqrt{C_2}}\vol_4,$$
  therefore this solution is a special case of the Freund-Rubin solution.

  In general, we can take
  $$B'=\frac{M}{r(r^6+M)}$$
  for some constant $M$ to solve (\ref{eqn:b}). We further get
  \bea
  e^{2A}=C_1\left(\frac{r^6}{r^6+M}\right)^{2/3},\quad
  e^{2B}=C_2\left(\frac{r^6}{r^6+M}\right)^{-1/3},\quad
  f'=\mp\frac{6MC_1^{3/2}r^5}{(r^6+M)^2}.\nonumber
  \eea
  For $M>0$, the corresponding $g_8$ is a complete metric on $\R^8\setminus\{0\}$, which is exactly the solution found in Duff-Stelle \cite{duff1991}. And for $M<0$, the corresponding $g_8$ is an incomplete metric on $\R^8\setminus B(\sqrt[6]{-M})$.
\medskip

\noindent{\bf (b). } We may also take $u=-4/r$ to solve (\ref{eqn:u}), in which case (\ref{eqn:ds3}) reduces to
  $$6(B')^2+\frac{24B'}{r}+\frac{48}{r^2}=\frac{e^{-6A}}{6}(f')^2,$$
  which implies that
  $$f'=\mp6e^{3A}\left(B'+\frac{2}{r}\right)=\pm(e^{3A})'.$$
  Combining it with (\ref{eqn:ds4}), we get
  \be
  B''+6(B')^2+\frac{19B'}{r}+\frac{12}{r^2}=0.\label{eqn:b'}
  \ee
  Two special solutions of (\ref{eqn:b'}) are $B'=-1/r$ and $B'=-2/r$, which correspond to $A'=-2/r$ and $A'=0$ respectively.

  In the first case, one can solve
  \bea
  e^{2A}=C_1r^{-4},\quad
  e^{2B}=C_2r^{-2},\quad
  f'=\pm\frac{6C_1^{3/2}}{r^7}\nonumber
  \eea
  to get a Freund-Rubin solution.

  In the second case one can similarly solve
  \bea
  e^A=C_1',\quad
  e^B=C_2'r^{-4},\quad
  f'=0.\nonumber
  \eea
  and again we get the Ricci-flat solution.

  The general solution to (\ref{eqn:b'}) is
  $$B'=-\frac{r^6+2M}{r(r^6+M)}.$$
  So we get
  \bea
  e^{2A}=C_1(r^6+M)^{-2/3},\quad
  e^{2B}=C_2r^{-4}(r^6+M)^{1/3},\quad
  f'=\pm\frac{6C_1^{3/2}r^5}{(r^6+M)^2},\nonumber
  \eea
  which is isomorphic to the Duff-Stelle solution. It is easy to verify that case (b) is related to case (a) by the inversion $r\leftrightarrow r^{-1}$.
\medskip

\noindent{\bf (c). } The general solution to (\ref{eqn:u}) is
  $$u=\frac{-4}{r(1-Cr^{12})}$$
  for some constant $C$, in which case $B$ satisfies
  $$B''+6(B')^2+\frac{19-7Cr^{12}}{r(1-Cr^{12})}B'+\frac{4(3-11Cr^{12})}{r^2(1-Cr^{12})^2}=0.$$
  To solve this Riccati equation, we first write $\rho=r^{12}$ and denote by $\dot{B}$ the expression
  $$\dot{B}=\frac{d B}{d\rho}=\frac{B'}{12r^{11}}.$$
  Therefore the above equation can be rewritten as
  $$\ddot{B}+6(\dot{B})^2+\frac{5-3C\rho}{2\rho(1-C\rho)}\dot{B}+\frac{3-11C\rho}{36\rho^2(1-C\rho)^2}=0.$$
  Let
  $$W=6\dot{B}+\frac{5-3C\rho}{4\rho(1-C\rho)},$$
  then $W$ satisfies the Riccati equation
  $$\dot{W}+W^2+\frac{9-118C\rho+9C^2\rho^2}{48\rho^2(1-C\rho)^2}=0.$$
  It is convenient to introduce the constant $\alpha=\sqrt{7/3}=1.5275\cdots$. By observation, one of the solution to this equation (when $C$ is positive) is given by
  $$W_0=\frac{-\frac{3}{4}C\rho+\alpha\sqrt{C\rho}+\frac{1}{4}}{\rho(1-C\rho)},$$
  and
  $$\int W_0 d\rho=\frac{1}{4}\left(\log\rho+2\log(1-C\rho)\right)+\alpha\log\left(\frac{1 +\sqrt{C\rho}}{1-\sqrt{C\rho}}\right).$$
  Therefore by general theory of Riccati equation, we know a general solution of above equation is of the form
  \bea
  W&=&W_0+\frac{\frac{1}{\sqrt{\rho}(1-C\rho)}\left(\frac{1-\sqrt{C\rho}}{1+\sqrt{C\rho}}\right)^{2\alpha}}
  {-\frac{1}{2\alpha\sqrt{C}}\left(\left(\frac{1-\sqrt{C\rho}}{1+\sqrt{C\rho}}\right)^{2\alpha}-1\right)+2M} \nonumber\\
  &=&\frac{-\frac{3}{4}C\rho+\alpha\sqrt{C\rho}+\frac{1}{4}}{\rho(1-C\rho)} +\frac{\frac{1}{\sqrt{\rho}(1-C\rho)}\left(\frac{1-\sqrt{C\rho}}{1+\sqrt{C\rho}}\right)^{2\alpha}}
  {-\frac{1}{2\alpha\sqrt{C}}\left(\left(\frac{1-\sqrt{C\rho}}{1+\sqrt{C\rho}}\right)^{2\alpha}-1\right)+2M}.\nonumber
  \eea
  It follows that
  $$\dot{B}=\frac{1}{6}\left(\frac{\alpha\sqrt{C\rho}-1}{\rho(1-C\rho)} +\frac{\frac{1}{\sqrt{\rho}(1-C\rho)}\left(\frac{1-\sqrt{C\rho}}{1+\sqrt{C\rho}}\right)^{2\alpha}}
  {-\frac{1}{2\alpha\sqrt{C}}\left(\left(\frac{1-\sqrt{C\rho}}{1+\sqrt{C\rho}}\right)^{2\alpha}-1\right)+2M}\right).$$
  Therefore we get
  $$e^{2B}=\frac{C_2}{r^4}\frac{\left(1+\sqrt{C}r^6\right)^{\frac{1}{3}(\alpha+1)}} {\left(1-\sqrt{C}r^6\right)^{\frac{1}{3}(\alpha-1)}}\left(M -\frac{1}{4\alpha\sqrt{C}}\left(\left(\frac{1-\sqrt{C}r^6}{1+\sqrt{C}r^6}\right)^{2\alpha}-1\right)\right) ^{1/3}.$$
  We can also solve that
  \bea
  &&e^{2A}=C_1\left(\frac{1-\sqrt{C}r^6}{1+\sqrt{C}r^6}\right)^{2\alpha/3}\left(M -\frac{1}{4\alpha\sqrt{C}}\left(\left(\frac{1-\sqrt{C}r^6}{1+\sqrt{C}r^6}\right)^{2\alpha}-1\right)\right) ^{-2/3},\nonumber\\
  &&f'=\pm \frac{6C_1^{3/2}\sqrt{1+4M\alpha\sqrt{C}}r^5}{\left(M -\frac{1}{4\alpha\sqrt{C}}\left( \left(\frac{1-\sqrt{C}r^6}{1+\sqrt{C}r^6}\right)^{2\alpha}-1\right)\right)^2}\frac{\left(1-\sqrt{C}r^6\right)^{2\alpha-1}}{\left(1+\sqrt{C}r^6\right)^{2\alpha+1}}.\nonumber
  \eea

  When $C<0$, one can similarly solve that
  $$\dot{B}=\frac{1}{6\rho(1-C\rho)}\left(\alpha\sqrt{-C\rho}\cot\left(\theta+2\alpha\arctan\left(\sqrt{-C\rho}\right)\right)-1\right),$$
  where $\theta$ is related to other constants by $\theta=2\alpha M\sqrt{-C}$.

  It follows that
  $$e^{2B}=C_2\left(\frac{1-Cr^{12}}{r^{12}}\frac{\sin\left(\theta+2\alpha\arctan\left(\sqrt{-C}r^6\right)\right)}{2\alpha\sqrt{-C}}\right)^{1/3}$$
  and
  \bea
  &&e^{2A}=C_1\left(\frac{\sin\left(\theta+2\alpha\arctan\left(\sqrt{-C}r^6\right)\right)}{2\alpha\sqrt{-C}}\right)^{-2/3},\nonumber\\
  &&f'=\frac{\mp56CC_1^{3/2}r^5}{(1-Cr^{12})\sin^2 \left(\theta+2\alpha\arctan\left(\sqrt{-C}r^6\right)\right)}.\nonumber
  \eea

By Theorem 3 we know that the only supersymmetric solutions in this 5-parameter family are 3(a) and 3(b), all isomorphic to the Duff-Stelle solution. Moreover, for any solution other than Duff-Stelle, the metric $(g_8)_{ij}=e^{2B}\delta_{ij}$ is incomplete.






\begin{appendix}

\section{(Multi-)Warped Product Metric}

Let $(M_i, g_i)$ be pseudo-Riemannian manifolds of dimension $n_i$ for $i=0,1,2,\dots,k$. A multi-warped product metric $g$ on the product manifold $M_0\times M_1\times\dots\times M_k$ is of the form
$$g=g_0+e^{f_1}g_1+\dots+e^{f_k}g_k,$$
where $f_1,\dots,f_k$ are smooth functions on $M_0$. In this appendix, we will compute the curvature tensors of (multi-)warped product metrics in terms of the curvature tensors $R_i$ of $(M_i,g_i)$. For simplicity of notation, we will use $X_j$ and $Y_j$ to denote vector fields tangential to $M_j$. These formulae can be found in literature (for instance \cite{dobarro2005}), we include them for the convenience of readers.

Using the Koszul identity, one can find that
\bea
&&\nabla_{X_0}Y_0=(\nabla_0)_{X_0}Y_0,\nonumber\nonumber \\
&&\nabla_{X_0}Y_j=\nabla_{Y_j}X_0=\frac{1}{2}X_0(f_j)Y_j,\nonumber\nonumber \\ &&\nabla_{X_j}Y_j=(\nabla_j)_{X_j}Y_j-\frac{1}{2}(X_j,Y_j)_{g_j}\nabla_0(e^{f_j}),\nonumber\nonumber \\
&&\nabla_{X_j}Y_l=0,\quad j\neq l.\nonumber
\eea
It follows that
\bea
&&R(X_0,Y_0,Z_0,W_0)=R_0(X_0,Y_0,Z_0,W_0),\nonumber\nonumber \\
&&R(X_0,Y_j,Z_0,W_j)=-e^{f_j}\left(\frac{1}{2}(\nabla_0^2f_j)(X_0,Z_0) +\frac{1}{4}X_0(f_j)Z_0(f_j)\right)(Y_j,W_j)_{g_j},\nonumber\nonumber \\
&&R(X_j,Y_j,Z_j,W_j)\nonumber\nonumber \\
&=&e^{f_j}R_j(X_j,Y_j,Z_j,W_j)+\frac{e^{2f_j}}{4}|\nabla_0f_j|^2\left((Y_j,Z_j)_{g_j}(X_j,W_j)_{g_j} -(X_j,Z_j)_{g_j}(Y_j,W_j)_{g_j}\right),\nonumber\nonumber \\ &&R(X_j,Y_l,Z_j,W_l)=-\frac{e^{f_j+f_l}}{4}(X_j,Z_j)_{g_j}(Y_l,W_l)_{g_l}(\nabla_0f_j,\nabla_0f_l), \quad j\neq l.\nonumber
\eea

It follows that the Ricci curvature of the warped metric $g$ is given by
\bea
&&\ric(X_0,Y_0)=\ric_0(X_0,Y_0)-\sum_{j=1}^kn_j\left(\frac{1}{2}(\nabla_0^2f_j)(X_0,Y_0)+ \frac{1}{4}X_0(f_j)Y_0(f_j)\right),\nonumber\nonumber \\
&&\ric(X_j,Y_j)=\ric_j(X_j,Y_j)
-\frac{e^{f_j}(X_j,Y_j)_{g_j}}{2}\left(\Delta_0(f_j)+\frac{1}{2}\sum_{l=1}^kn_l(\nabla_0 f_j,\nabla_0f_l)\right),\nonumber
\eea
and all other components are zero.

Similarly the scalar curvature can be computed as
$$S=S_0+\sum_{j=1}^ke^{-f_j}S_j-\sum_{j=1}^kn_j\left(\Delta_0(f_j)+\frac{1}{4}|\nabla_0f_j|^2\right) -\frac{1}{4}\left|\sum_{j=1}^ln_j\nabla_0 f_j\right|^2.$$

\end{appendix}

\bigskip
Department of Mathematics, Columbia University, New York, NY 10027, USA

\smallskip
tfei@math.columbia.edu, bguo@math.columbia.edu, phong@math.columbia.edu


\begin{thebibliography}{99}

{\small

\bibitem{baum1991} Baum, H., Th. Friedrich, R. Grunewald, and I. Kath, {\em Twistors and Killing Spinors on Riemannian Manifolds}, Teubner-Texte zur Mathematik 124, B.G. Teubner, 1991.

\bibitem{B} Beauville, A., {\em Some remarks on K\"ahler manifolds with $c_1 = 0$}, Classification of algebraic and analytic manifolds (Katata, 1982), 1 -- 26, Progr. Math. 39, Birkh\"auser, Boston, MA, 1983.

\bibitem{CL} Cheng, S.-Y. and P.W.-K. Li, {\em Heat kernel estimates and lower bounds of eigenvalues}, Comm. Math. Helvetici 56 (1981), 327 -- 338.

\bibitem{Cl} Clancy, R., {\em New examples of compact manifolds with holonomy $\textrm{Spin}(7)$}, Ann. Global Anal. Geom. 40 (2011), 203 -- 222.

\bibitem{CH1} Conlon, R., and H.-J. Hein, {\em Asymptotically conical Calabi-Yau manifolds, I}, Duke Math. J. 162 (2013), 2855 -- 2902.

\bibitem{CH} Conlon, R., and H.-J. Hein, {\em Asymptotically conical Calabi-Yau metrics on quasi-projective varieties}, Geom. Funct. Anal. 25 (2015), 517 -- 552.

\bibitem{CR} Conlon, R., and F. Rochon, {\em New examples of complete Calabi-Yau metrics on ${\mathbb C}^n$ for $n\ge 3$}. arXiv: 1705.08788.

\bibitem{cremmer1978} Cremmer, E., B. Julia, and J. Scherk, {\em Supergravity theory in 11 dimensions}, Phys. Lett. B 76 (1978), 409 -- 412.

\bibitem{DH} D'Hoker, E., {\em Exact M-Theory Solutions, Integrable Systems, and Superalgebras}, SIGMA 11 (2015), 609 -- 628.

\bibitem{dW1} de Wit, B. and H. Nicolai, {\em The consistency of the $S^{7}$ truncation in $d=11$ supergravity}, Nucl. Phys. B 281 (1986), 211 -- 240.

\bibitem{dW2} de Wit, B., D. Smit, and N.D. Hari Dass, {\em Residual supersymmetry of compactified $d=10$ supergravity}, Nucl. Phys. B 283 (1987), 165 -- 191.

\bibitem{dobarro2005} Dobarro, F. and B. \"Unal, {\em Curvature of multiply warped products}, J. Geom. Phys. 55 (2005), 75 -- 106.

\bibitem{duff} Duff, M.J., {\em The world in eleven dimensions: supergravity, supermembranes and M-theory}, Studies in High Energy Physics and Cosmology, IOP Publishing, Taylor \& Francis, 1999.

\bibitem{duff1991} Duff, M.J. and K.S. Stelle, {\em Multi-membrane solutions of {D}=11 supergravity}, Phys. Lett. B 253 (1991), 113 -- 118.

\bibitem{englert1982} Englert, F., {\em Spontaneous compactification of eleven-dimensional supergravity}, Phys. Lett. B 119 (1982), 339 -- 342.

\bibitem{Fig} Figueroa-O'Farrill, J. and G. Papadopoulos, {\em Maximally supersymmetric solutions of ten and eleven-dimensional supergravities}, JHEP 048 (2003), 25 pp.

\bibitem{freund1980} Freund, P.G.O. and M.A. Rubin, {\em Dynamics of dimensional reduction}, Phys. Lett. B 97 (1980), 233 -- 235.

\bibitem{friedrich1997} Friedrich, Th., I. Kath, A. Moroianu, and U. Semmelmann, {\em On nearly parallel ${G}_2$-structures}, J. Geom. Phys. 23 (1997), 259 -- 286.

\bibitem{GP} Gauntlett, J. and S. Pakis, {\em The geometry of $D=11$ Killing spinors}, JHEP 039 (2003), 32 pp.

\bibitem{HW} Horava, P. and E. Witten, {\em Heterotic and type I string dynamics from eleven dimensions}, Nucl. Phys. B 460 (1996), 506 -- 524.

\bibitem{H1} Hull, C.M., {\em Supersymmetry with torsion and space-time supersymmetry}, 1st Torino Meeting on Superunification and Extra Dimensions, 347 -- 375, World Scientific, 1986.

\bibitem{H2} Hull, C.M., {\em Holonomy and symmetry in M-branes}, hep-th/0305039.

\bibitem{Jo} Joyce, D., {\em Compact Manifolds with Special Holonomy}, Oxford Math. Monogr., Oxford University Press, Oxford, 2000.

\bibitem{LY} Li, P.W.-K. and Yau, S.-T., {\em On the parabolic kernel of the Schr\"odinger operator}, Acta Math. 156 (1986), 153 -- 201.

\bibitem{Li} Li, Y., {\em A new complete Calabi-Yau metric on ${\mathbb C}^3$}, arXiv: 1705.07026.

\bibitem{moroianu2000} Moroianu, A. and U. Semmelmann, {\em Parallel spinors and holonomy groups}, J. Math. Phys. 41 (2000), 2395 -- 2402.

\bibitem{PvN} Pope, C. and P. van Nieuwenhuizen, {\em Compactifications of $d=11$ supergravity on K\"ahler manifolds}, Commun. Math. Phys. 122 (1989), 281 -- 292.

\bibitem{PW} Pope, C. and N. Warner, {\em An $SU(4)$ invariant compactification of $d=11$  supergravity on a stretched seven-sphere}, Phys. Lett. B 150, 352 -- 356.

\bibitem{S} Strominger, A.E., {\em Superstrings with torsion}, Nuclear Physics B 274 (1986), 253 -- 284.

\bibitem{Sz} Sz\'ekelyhidi, G., {\em Degenerations of ${\mathbf C}^n$ and Calabi-Yau metrics}, arXiv: 1706.00357.

\bibitem{Ta} Taylor, C.J.-Y., {\em New examples of compact 8-manifolds of holonomy $\textrm{Spin}(7)$}, Math. Res. Lett. 6 (1999), 557 -- 561.

\bibitem{TY} Tian, G. and Yau, S.-T., {\em Complete K\"ahler manifolds with zero Ricci curvature, II}, Invent. Math. 106 (1991), 27 -- 60.

\bibitem{T} Townsend, P., {\em The eleven-dimensional supermembrane revisited}, Phys. Lett. B 350 (1995), 184 -- 188.

\bibitem{W} Witten, E., {\em String theory dynamics in various dimensions}, Nucl. Phys. B 443 (1995), 85 -- 126.

\bibitem{wang1989} Wang, M.Y.-K., {\em Parallel spinors and parallel forms},  Ann. Global Anal. Geom. 7 (1989), 59 -- 68.

\bibitem{Y} Yau, S.-T., {\em On the Ricci curvature of a compact K\"ahler manifold and the complex Monge-Amp\`ere equation. I.} Comm. Pure Appl. Math. 31 (1978), 339 -- 411.
}

\end{thebibliography}
\end{document}